\documentclass[12pt]{article}
\usepackage{info-ineq}
\begin{document}
\title{Some Quantum Information Inequalities\\
from a Quantum Bayesian Networks Perspective}

\author{Robert R. Tucci\\
        P.O. Box 226\\
        Bedford,  MA   01730\\
        tucci@ar-tiste.com}

\date{\today}
\maketitle
\vskip2cm
\section*{Abstract}
This is primarily a pedagogical paper.
The paper re-visits
some well-known quantum
information theory
inequalities.
It does this from a
quantum Bayesian networks
perspective. The paper
illustrates some of the benefits
of using quantum Bayesian networks
to discuss quantum SIT (Shannon Information Theory).

\newpage

\section{Introduction}
For a good textbook on classical (non-quantum)
Shannon
Information Theory (SIT), see, for example,
Ref.\cite{CovTh}
by Cover and Thomas.
For a good
textbook
on quantum SIT, see, for example,
Ref.\cite{Wilde} by Wilde.

This paper is written assuming that
the reader has first read
a previous paper by the same
author, Ref.\cite{Tuc-mixology},
which is an introduction
to quantum Bayesian networks for mixed states.

This paper re-visits
some well-known quantum
information theory
inequalities (mostly
the monotonicity
of the relative entropy
and consequences thereof).
It does this from a
quantum Bayesian networks
perspective. The paper
illustrates some of the benefits
of using quantum Bayesian networks
to discuss quantum SIT.

\section{Preliminaries and Notation}

Reading all of Ref.\cite{Tuc-mixology}
is a
prerequisite to reading this paper.
This
section
will introduce
only notation
which  hasn't
been defined already in
Ref.\cite{Tuc-mixology}.

Let
\beq
S_{\rvx,\rvy} =
S_\rvx\times S_\rvy=
\{(x,y):x\in S_\rvx, y\in S_\rvy\}
\;,
\eeq
\beq
\calh_{\rvx,\rvy} =
\calh_\rvx\otimes \calh_\rvy=
span\{\ket{x}_\rvx\ket{y}_\rvy:
x\in S_\rvx, y\in S_\rvy\}
\;.
\eeq

Suppose
$\{P_{\rvx,\rvy}(x,y)\}_{\forall x,y}
\in pd(S_{\rvx,\rvy})$.
We will often use
the expectation operators
$E_x = \sum_x P(x)$,
 $E_{x,y}=\sum_{x,y} P(x,y)$,
 and
$E_{y|x}=\sum_y P(y|x)$.
Note that $E_{x,y} = E_x E_{y|x}$.
Let
\beq
P(x:y) = \frac{P(x,y)}{P(x)P(y)}
\;.
\eeq
Note that
$E_{x} P(x:y) = E_{y} P(x:y)=1$.

We will use the following
measures of various
types of information (entropy):

\begin{itemize}
\item
The (plain) entropy of the
random variable $\rvx$ is defined
in the classical case by

\beq
H(\rvx) =
E_x \ln \frac{1}{P(x)}
\;,
\eeq
which we also call
$H_{P_\rvx}(\rvx)$,
$H\{P(x)\}_{\forall x}$,
and
$H(P_\rvx)$.
This quantity measures the
spread of $P_\rvx$.
The quantum generalization of this is, for
$\rho_{\rvx}\in
dm(\calh_{\rvx})$,

\beq
S(\rvx) =
-\tr_\rvx (\rho_\rvx\ln \rho_\rvx)
\;,
\eeq
which we also call
$S_{\rho_\rvx}(\rvx)$ and
$S(\rho_\rvx)$.

One can also consider
plain entropy for
a joint random variable
$\rvx=(\rvx_1,\rvx_2)$.
In the
classical
case,
for $P_{\rvx_1,\rvx_2}\in pd(S_{\rvx_1,\rvx_2})$
with marginal probability distributions $P_{\rvx_1}$
and $P_{\rvx_2}$,
one defines a joint entropy $H(\rvx_1,\rvx_2)=H(\rvx)$
and partial entropies
$H(\rvx_1)$ and $H(\rvx_2)$.
The quantum generalization of this is, for
$\rho_{\rvx_1,\rvx_2}\in dm(\calh_{\rvx_1,\rvx_2})$
with partial density matrices
$\rho_{\rvx_1}$ and $\rho_{\rvx_2}$,
a joint entropy
$S(\rvx_1,\rvx_2)=S(\rvx)$
with partial entropies
$S(\rvx_1)$ and $S(\rvx_2)$.

\item
The conditional entropy of $\rvy$ given $\rvx$
is defined
in the classical case by

\beqa
H(\rvy|\rvx) &=&
E_{x,y} \ln \frac{1}{P(y|x)}
\\
&=&
H(\rvy,\rvx)-H(\rvx)
\;,
\eeqa
which we also call
$H_{P_{\rvx, \rvy}}(\rvy|\rvx)$.
This quantity measures  the conditional
 spread
of $\rvy$ given $\rvx$.
The quantum generalization of this is, for
$\rho_{\rvx,\rvy}\in
dm(\calh_{\rvx,\rvy})$,

\beq
S(\rvy|\rvx) =
S(\rvy,\rvx) -S(\rvx)
\;,
\eeq
which we also call
$S_{\rho_{\rvx, \rvy}}(\rvy|\rvx)$.

\item The Mutual Information (MI)
of $\rvx$ and $\rvy$
is defined
in the classical case by

\beqa
H(\rvy:\rvx) &=&
E_{x,y} \ln
P(x:y)= E_x E_y P(x:y) \ln P(x:y)
\\
&=&
H(\rvx) + H(\rvy) - H(\rvy,\rvx)
\;,
\eeqa
which we also call
$H_{P_{\rvx,\rvy}}(\rvy:\rvx)$.
This quantity measures the correlation
between $\rvx$ and $\rvy$.
The quantum generalization of this is, for
$\rho_{\rvx,\rvy}\in
dm(\calh_{\rvx,\rvy})$,

\beq
S(\rvy:\rvx) =
S(\rvx) + S(\rvy) - S(\rvx,\rvy)
\;,
\eeq
which we also call
$S_{\rho_{\rvx,\rvy}}(\rvy:\rvx)$.

\item The Conditional Mutual Information (CMI,
which
can be read as ``see me")
of $\rvx$ and $\rvy$
given $\rv{\lam}$
is defined
in the classical case by:

\beqa
H(\rvy:\rvx|\rv{\lam})
&=&
E_{x,y,\lam} \ln
\frac{P(x,y|\lam)}{P(x|\lam)P(y|\lam)}
\\
&=&
E_{x,y,\lam} \ln
\frac{P(x,y,\lam)P(\lam)}{P(x,\lam)P(y,\lam)}
\\
&=&
H(\rvx|\rv{\lam}) + H(\rvy|\rv{\lam})
- H(\rvy,\rvx|\rv{\lam})
\;,
\eeqa
which we also call
$H_{P_{\rvx,\rvy,\rv{\lam}}}(\rvy:\rvx|\rv{\lam})$.
This
quantity measures the conditional correlation
of $\rvx$ and $\rvy$ given $\rv{\lam}$.
The quantum generalization of this is, for
$\rho_{\rvx,\rvy,\rv{\lam}}\in
dm(\calh_{\rvx,\rvy,\rv{\lam}})$,

\beq
S(\rvy:\rvx|\rv{\lam})
=
S(\rvx|\rv{\lam}) + S(\rvy|\rv{\lam})
- S(\rvy,\rvx|\rv{\lam})
\;,
\eeq
which we also call
$S_{\rho_{\rvx,\rvy,\rv{\lam}}}(\rvy:\rvx|\rv{\lam})$

\item The relative
information
of $P\in pd(S_\rvx)$
divided by $Q\in pd(S_\rvx)$
is defined by

\beq
D\{P(x)//Q(x)\}_{\forall x} =
\sum_x P(x)\ln\frac{P(x)}{Q(x)}
\;,
\eeq
which we also call
$D(P_{\rvx}//Q_\rvx)$.
The quantum generalization of this is, for
$\rho_{\rvx}, \sigma_\rvx\in
dm(\calh_{\rvx})$,

\beq
D(\rho_\rvx//\sigma_\rvx) =
\tr_\rvx\left(\rho_\rvx (\ln \rho_\rvx
-\ln \sigma_\rvx)\right)
\;.
\eeq

\end{itemize}

Note that we
define entropies
using natural logs. Our
strategy is to
use natural log entropies
for all intermediate analytical
calculations, and to
convert
to base-2 logs
at the end of those
calculations if
a base-2 log numerical answer
is desired. Such a conversion is
of course trivial
using $\log_2 x = \frac{\ln x}{\ln 2}$ and
$\ln 2 = 0.6931$

The notation $\atrho{\rho}\calf\atrhoend$
will be used to
indicate that all quantum
entropies $S(\cdot)$
in statement $\calf$ are to be evaluated
at density matrix $\rho$.
For example,
$\atrho{\rho}
S(\rva) + S(\rvb|\rvc)=0
\atrhoend$
will stand for
$S_\rho(\rva) + S_\rho(\rvb|\rvc)=0$.

Define
\beq
I_\rvx = \sum_{x\in S_\rvx}
\ket{x}_\rvx\bra{x}_\rvx
\;.
\eeq
Define $1^{N}$ to be the $N$-tuple
whose $N$ components are all equal to one.

Recall
from Ref.\cite{Tuc-mixology}
that
an amplitude
$\{A(y|x)\}_{\forall y,x}$ is
said to be an {\bf isometry} if

\beq
\sum_{y}
\sandb{A(y|x)}
\sandb{
\hc\\
x\rarrow x'
}
=
\delta_x^{x'}
\;
\eeq
for all $x,x'\in S_\rvx$.

\section{Monotonicity of Relative
Entropy (MRE)}
In this section, we will state
the monotonicity of
 the relative entropy (MRE,
which can be read as ``more") and
derive some of its many consequences,
such as $MI\geq 0$,
$CMI\geq 0$,
and the data processing inequalities.

\subsection{General MRE Inequality}

\begin{claim}
Suppose $\{P_\rva(a)\}_{\forall a\in S_\rva}$
and $\{Q_\rva(a)\}_{\forall a\in S_\rva}$
are both probability distributions.
Suppose $\{T_{\rvb|\rva}(b|a)\}_
{\forall (b,a)\in S_{\rvb,\rva}}$ is a
transition probability matrix, meaning that
its entries are non-negative and
satisfy $\sum_b T_{\rvb|\rva}(b|a)= 1$
for any $a\in S_\rva$.
Then

\beq
D(T_{\rvb|\rva}P_\rva//
T_{\rvb|\rva}Q_\rva)
\leq
D(P_\rva//Q_\rva)
\;,\label{eq-mono-cla}
\eeq
where we are overloading the symbol
$T_{\rvb|\rva}$ so that it
 stands also for an $N_\rvb\times N_\rva$
 matrix,
and we are overloading the symbols
$P_\rva, Q_\rva$
so that they stand also for
$N_\rva$-dimensional column vectors.
\end{claim}
\proof

\beqa
D(P//Q) &=&
\sum_a P(a) \ln \frac{P(a)}{Q(a)}
\\
&=&
\sum_{a,b} T(b|a)P(a) \ln \frac{T(b|a)P(a)}{T(b|a)Q(a)}
\\
&\geq&
\sum_{b} (TP)(b) \ln \frac{(TP)(b)}{(TQ)(b)}
\label{eq-from-log-sum}
\\
&=&
D(TP//TQ)
\;
\eeqa
Eq.(\ref{eq-from-log-sum}) follows
from the so called log-sum inequality (See
Ref.\cite{CovTh}).
\qed

Recall from Ref.\cite{Tuc-mixology}
that a channel superoperator
$\calt_{\rvb|\rva}$
is a map from $dm(\calh_\rva)$
to $dm(\calh_\rvb)$ which can be expressed
as

\beq
\calt_{\rvb|\rva}(\cdot)=
\sum_\mu K_\mu (\cdot) K^\dagger_\mu
\;,
\eeq
where the operators $K_\mu:\calh_\rva\rarrow \calh_\rvb$,
called Krauss operators, satisfy:

\beq
\sum_\mu K^\dagger_\mu K_\mu = 1
\;.
\eeq
Ref.\cite{Tuc-mixology}
explains how a channel
superop can be portrayed
in terms of QB nets as
a two body
scattering diagram.

\begin{claim}
Suppose $\rho_\rva, \sigma_\rva\in dm(\calh_\rva)$
and $\calt_{\rvb|\rva}:dm(\calh_\rva)\rarrow dm(\calh_\rvb)$
is a channel superop.
Then
\beq
D(\calt_{\rvb|\rva}(\rho_\rva)//
\calt_{\rvb|\rva}(\sigma_\rva))
\leq
D(\rho_\rva//\sigma_\rva)
\;\label{eq-mono-qua}
\eeq
\end{claim}
\proof
See Ref.\cite{Wilde}
and original references therein.
\qed

Note that
Eq.(\ref{eq-mono-cla})
is a special case of
Eq.(\ref{eq-mono-qua}).
Indeed,
if $\calt_{\rvb|\rva}$
has Krauss operators $\{K_\mu\}_{\forall \mu}$,
then let

\begin{subequations}
\label{eq-rb-k-ra-k}
\beq
\rho_\rvb =
\sum_\mu K_\mu \rho_\rva K^\dagger_\mu
\;,
\eeq

\beq
\sigma_\rvb =
\sum_\mu K_\mu \sigma_\rva K^\dagger_\mu
\;.
\eeq
\end{subequations}
Assume $\rho_\rva$ and $\sigma_\rva$
can both be diagonalized in the same basis
as follows

\begin{subequations}
\beq
\rho_\rva =
\sum_a \sandb{\ket{a}_\rva}
P_\rva(a)\sandb{\hc}
\;,
\eeq

\beq
\sigma_\rva =
\sum_a \sandb{\ket{a}_\rva}
Q_\rva(a)\sandb{\hc}
\;.
\eeq
\label{eq-diag-same-basis}\end{subequations}
Likewise, assume that
$\rho_\rvb$ and $\sigma_\rvb$
can both be diagonalized in the same
basis. Thus assume Eqs.(\ref{eq-diag-same-basis}),
but with the letters $a$'s replaced by $b$'s.
Then Eqs.(\ref{eq-rb-k-ra-k}) reduce to

\begin{subequations}
\beq
P_\rvb = T_{\rvb|\rva} P_\rva
\;,
\eeq

\beq
Q_\rvb = T_{\rvb|\rva} Q_\rva
\;,
\eeq
\end{subequations}
where

\beq
T_{\rvb|\rva}(b|a)=
\sum_\mu |\av{b|K_\mu|a}|^2
\;\label{eq-t-krauss}
\eeq
for all $a\in S_\rva$
and $b\in S_\rvb$.
Clearly, this $T_{\rvb|\rva}$ satisfies
$\sum_b T(b|a)=1$.
Therefore
the quantum MRE
with diagonal density matrices
is just the classical MRE.

\subsection{Subadditivity of Joint Entropy (MI$\geq$0)}

For any random variables $\rva,\rvb$,
\beq
H(\rva, \rvb)\leq H(\rva) + H(\rvb)
\;.
\eeq
This is sometimes called
the subadditivity of the joint
entropy, or the independence upper bound
on the joint entropy.
It can also be written as
(i.e., conditioning reduces entropy)

\beq
H(\rvb|\rva)\leq H(\rvb)
\;,
\eeq
or as (MI$\geq 0$)

\beq
H(\rvb:\rva)\geq 0
\;.
\eeq

\begin{claim} (MI $\geq 0$)
For any $\rho\in dm(\calh_{\rva,\rvb})$,
\beq
S(\rva, \rvb)\leq S(\rva) + S(\rvb)
\;,
\eeq
or, equivalently,

\beq
S(\rvb|\rva)\leq S(\rvb)
\;,
\eeq
or, equivalently,

\beq
S(\rvb:\rva)\geq 0
\;.
\eeq
\end{claim}
\proof
Apply MRE
with
$\calt = \tr_{\rva}$.

\beq
0=D(\rho_\rvb//\rho_\rvb)
\leq
D(\rho_{\rva,\rvb}//
\rho_{\rva}
\rho_{\rvb})=S(\rva:\rvb)
\;.
\eeq
\qed

\subsection{Strong Subadditivity of
Joint Entropy (CMI$\geq$0)}

For any random variables $\rva,\rvb,\rve$,
\beq
H(\rva, \rvb|\rve)\leq H(\rva|\rve) + H(\rvb|\rve)
\;.
\eeq
This is sometimes called
the strong subadditivity of the joint
entropy.
It can also be written as

\beq
H(\rvb|\rva,\rve)\leq H(\rvb|\rve)
\;,
\eeq
or as (CMI $\geq 0$)

\beq
H(\rvb:\rva|\rve)\geq 0
\;.
\eeq

\begin{claim} (CMI $\geq 0$)
For any $\rho\in dm(\calh_{\rva,\rvb,\rve})$,
\beq
S(\rva, \rvb|\rve)\leq S(\rva|\rve) + S(\rvb|\rve)
\;,
\eeq
or, equivalently,

\beq
S(\rvb|\rva,\rve)\leq S(\rvb|\rve)
\;,
\eeq
or, equivalently,

\beq
S(\rvb:\rva|\rve)\geq 0
\;.
\eeq
\end{claim}
\proof
Apply MRE
with
$\calt = \tr_{\rva}$
to get

\beq
D(\rho_{\rvb,\rve}//
\rho_{\rve}\frac{I_\rvb}{N_\rvb}
)
\leq
D(
\rho_{\rva,\rvb,\rve}//
\rho_{\rva,\rve}\frac{I_\rvb}{N_\rvb})
\;.
\eeq
Then note that

\beq
S(\rva:\rvb|\rve)=
D(
\rho_{\rva,\rvb,\rve}//
\rho_{\rva,\rve}\frac{I_\rvb}{N_\rvb})
-
D(\rho_{\rvb,\rve}//
\rho_{\rve}\frac{I_\rvb}{N_\rvb}
)
\;.
\eeq
\qed

\subsection{Araki-Lieb Inequality}

\begin{claim}(Araki-Lieb Inequality\cite{ArLi})
For any $\rho\in dm(\calh_{\rva,\rvb})$,

\beq
|S(\rva)-S(\rvb)|\leq S(\rva,\rvb)
\;.
\eeq
or, equivalently,

\beq
\left\{
\begin{array}{l}
-S(\rvb)\leq S(\rvb|\rva)\\
-S(\rva)\leq S(\rva|\rvb)
\end{array}
\right.
\;.
\eeq
\end{claim}
\proof
Consider a pure state $\rho_{\rva,\rvb,\rve}
\in dm(\calh_{\rva,\rvb,\rve})$
with partial trace $\rho_{\rva,\rvb}$.
Then

\beq
S(\rvb,\rve)\leq S(\rvb) + S(\rve)
\;.
\label{eq-be-st-ad}
\eeq
According to Claim \ref{claim-sa-is-sb},
$S(\rvb,\rve)=S(\rva)$ and $S(\rve)=S(\rva,\rvb)$.
These two
identities allow us to excise any
mention of $\rve$ from Eq.(\ref{eq-be-st-ad}).
Thus
Eq.(\ref{eq-be-st-ad}) is equivalent to

\beq
S(\rva)\leq S(\rvb) + S(\rva,\rvb)
\;,
\eeq
which immediately gives

\beq
-S(\rvb)\leq S(\rvb|\rva)
\;.
\eeq
\qed

Note that classically, one has
\beq
0
\stackrel{(a)}{\leq}
H(\rvb|\rva)
\stackrel{(b)}{\leq}
H(\rvb)
\;.
\eeq
Inequality (a) follows
from the definition of
$H(\rvb|\rva)$, and (b)
follows from MI$\geq 0$.

For quantum states, on the other hand,

\beq
-S(\rvb)
\stackrel{(a)}{\leq}
S(\rvb|\rva)
\stackrel{(b)}{\leq}
S(\rvb)
\;,
\eeq
or, equivalently,

\beq
|S(\rvb|\rva)|\leq S(\rvb)
\;.
\eeq
Inequality (a) follows
from the Araki-Lieb inequality, and (b)
follows from MI$\geq 0$.

\subsection{Monotonicity (Only in Some Special Cases) of Plain Entropy }

Consider the two node CB net
\beq
\entrymodifiers={++[o][F-]}
\xymatrix{
\rvb&\rva\ar[l]
}
\;.\label{eq-ab-cbnet}
\eeq
For this net, $P_\rvb = P_{\rvb|\rva}P_\rva$.
Assume also that
$P_{\rvb|\rva}$ is a square matrix (i.e.,
that $N_\rva=N_\rvb$)
and
that it is doubly stochastic (i.e.,
that $\sum_b P(b|a)=1$
for all $a$, and $\sum_a P(b|a)=1$
for all $b$. In other words,
each of
its columns and rows sums to one.).
Then
the classical MRE
implies

\beq
D(P_\rvb//\frac{1^{N}}{N})
\leq
D(P_\rva//\frac{1^{N}}{N})
\;,
\eeq
where $N=N_\rva=N_\rvb$.
(The reason we need $\sum _a P(b|a)=1$
 is that we
must have $P_{\rvb|\rva}1^N = 1^N$).
Next note that for any
random variable $\rvx$,

\beq
H(P_\rvx) = \ln(N_\rvx) -
D(P_\rvx//\frac{1^{N_
\rvx}}{N_\rvx})
\;.
\eeq
Thus,

\beq
H(\rvb)\geq H(\rva)
\;.
\eeq
Thus,
when $P_{\rvb|\rva}$
is square and doubly stochastic,
$P_{\rvb}$
has a larger spread
than $P_\rva$.
This situation is sometimes
described
by saying that ``mixing" increases
entropy.

An important scenario
where the opposite is
the case and
$P_{\rvb}$
has a {\it smaller} spread
than $P_{\rva}$
is
when
$\rvb=f(\rva)$
for some deterministic
function $f:S_\rva\rarrow S_\rvb$.
In this case, $P(b|a)=\delta(b, f(a))$
(clearly not a doubly stochastic
transition matrix). Thus

\beq
H(\rvb,\rva)= H(\rvb|\rva)+H(\rva)=H(\rva)
\;
\eeq
Also
\beq
H(\rvb,\rva)= H(\rva|\rvb)+H(\rvb)
\;
\eeq
Hence

\beq
H(\rva|\rvb)+H(\rvb) = H(\rva)
\;
\eeq
But $H(\rva|\rvb)\geq 0$. Hence

\beq
H(\rvb)=H(f(\rva))\leq H(\rva)
\;.
\eeq
Loosely speaking,
the random variable $f(\rva)$
varies over a smaller range
than $\rva$ (unless $f()$ is a
bijection),
so $f(\rva)$ has a smaller spread than $\rva$.

\begin{claim}\label{cl-more-plain-h}
Suppose $\rho_\rva\in dm(\calh_\rva)$
and $\calt_{\rvb|\rva}:dm(\calh_\rva)
\rarrow dm(\calh_\rvb)$
is a square (i.e., $N_\rva=N_\rvb$)
channel superop
such that $I_\rvb=\calt_{\rvb|\rva}(I_\rva)$. Then

\beq
S(\calt_{\rvb|\rva}(\rho_\rva))\geq
S(\rho_\rva)
\;.
\eeq
\end{claim}
\proof
Let
$\rho_\rvb = \calt_{\rvb|\rva}(\rho_\rva)$.
Then MRE implies

\beq
D(\rho_\rvb//\frac{I_\rvb}{N})
\leq
D(\rho_\rva//\frac{I_\rvb}{N})
\;,
\eeq
where $N=N_\rva=N_\rvb$.
Now note that for $\rvx=\rva,\rvb$,

\beq
S(\rho_\rvx) = \ln(N_\rvx) -
D(\rho_\rvx//\frac{I_\rvx}{N_\rvx})
\;.
\eeq
\qed

\subsection{Entropy of Measurement}
\label{sec-ent-mea}

Applying the cl$(\cdot)$
operator to a node (``classicizing" it)
is like a ``measurement".
Thus, the following inequality is
often called the entropy of measurement
inequality.

\begin{claim}
For any $\rho_\rvx \in \calh_\rvx$
and orthonormal basis
$\{\ket{x}_\rvx\}_{\forall x}$,

\beq
H\{\av{x|\rho_\rvx|x}\}_{\forall x}
\geq S(\rho_\rvx)
\;,
\eeq
or, equivalently,
\beq
S_{\rho_{\rvx_{cl}}}(\rvx_{cl})
\geq
S_{\rho_{\rvx}}(\rvx)
\;.
\eeq
\end{claim}
\proof: This is a special case of Claim
\ref{cl-more-plain-h} with $\rva = \rvx$,
$\rvb=\rvx_{cl}$,
and $\calt = {\rm cl}_\rvx$.
\qed

Note that one can
prove many other similar inequalities
by appealing to
MRE with $\calt = {\rm cl}_\rvx$.
For instance,
for any $\rho_{\rva,\rvb}\in \calh_{\rva,\rvb}$,

\beq
S_{\rho_{\rvb,\rva_{cl}}}(\rvb,\rva_{cl})
\geq
S_{\rho_{\rvb,\rva}}(\rvb,\rva)
\;,
\eeq
and

\beq
S_{\rho_{\rvb,\rva_{cl}}}(\rvb:\rva_{cl})
\leq
S_{\rho_{\rvb,\rva}}(\rvb:\rva)
\;.
\eeq

\subsection{Entropy of Preparation}

An ensemble
$\{\sqrt{w_j}\ket{\psi_j}\}_{\forall j}$
for a system can
be described as a
preparation of the system.
Thus the following inequality is
often called the entropy
of preparation inequality.

\begin{claim}
Suppose the weights $\{w_j\}_{\forall j\in S_\rvj}$
are non-negative numbers that sum to one,
and $\{\ket{\psi_j}_\rvx\}_{\forall j\in S_\rvj}$
are normalized
states that span $\calh_\rvx$. Let
\beq
\rho_\rvx = \sum_j w_j
\sandb{\ket{\psi_j}_\rvx}\sandb{\hc}
=
\tr_{\rvj}(\rho_{\rvx,\rvj})
\;
\eeq
where $\rho_{\rvx,\rvj}\in
dm(\calh_{\rvx,\rvj})$
is a pure state (a ``purification"
 of $\rho_\rvx$).
Then

\beq
H\{w_j\}_{\forall j} \geq S(\rho_\rvx)
\;,
\eeq
or, equivalently,

\beq
S_{\rho_{\rvx,\rvj_{cl}}}(\rvj_{cl})
\geq S_{\rho_{\rvx}}(\rvx)
\;.
\eeq
The inequality becomes an equality
iff the states
$\{\ket{\psi_j}_\rvx\}_{\forall j}$
are orthonormal, in which case the weights
$\{w_j\}_{\forall j}$
are the eigenvalues of $\rho_\rvx$.
\end{claim}
\proof
Let
\beq
\rho_{\rvx,\rvj}=
\sandb{\ket{\psi}_{\rvx,\rvj}}
\sandb{\hc}
\;,
\eeq
where

\beq
\ket{\psi}_{\rvx,\rvj} =
\sandb{\sum_{x,j}A(x,j)
\begin{array}{c}
\ket{x}_\rvx\\
\ket{j}_\rvj
\end{array}
}
=
\sandb{
\entrymodifiers={++[o][F-]}
\xymatrix{\rvx&\rvj\ar[l]}
}
\;,
\eeq
where

\beq
A(x,j) = A(x|j)A(j)
\;,
\eeq
and

\beq
A(x|j) = \av{x|\psi_j}
,\;\;\;
A(j) = \sqrt{w_j}
\;.
\eeq
Then

\beq
S_{\rho_{\rvx,\rvj_{cl}}}(\rvj_{cl})
\stackrel{(a)}{\geq}
S_{\rho_{\rvx,\rvj}}(\rvj)
\stackrel{(b)}{=}
S_{\rho_{\rvx,\rvj}}(\rvx)
=
S_{\rho_{\rvx}}(\rvx)
\;
\eeq

$(a)$ follows from the entropy
of measurement inequality (Section \ref{sec-ent-mea}).
Note that $(a)$ becomes an equality
iff the states $\{\ket{\psi_j}_\rvx\}_{\forall j}$
are orthonormal.

$(b)$ follows because $\rho_{\rvx,\rvj}$
is a pure state.
\qed

\subsection{Data Processing Inequalities}

Consider the following CB net
\beq
\entrymodifiers={++[o][F-]}
\xymatrix{
\rvc&\rvb\ar[l]&\rva\ar[l]
}
\;.
\eeq
Classical MRE with $T=P_{\rvc|\rvb}$
implies

\beq
D(P_{\rvc,\rva}//P_\rvc P_\rva)
\leq
D(P_{\rvb,\rva}//P_\rvb P_\rva)
\;.
\eeq
Thus

\beq
H(\rvc:\rva)\leq H(\rvb:\rva)
\;.\label{eq-dp-classical}
\eeq
Eq.(\ref{eq-dp-classical})
is called a data processing
inequality.

Next consider the following CB net
\beq
\entrymodifiers={++++[o][F-]}
\xymatrix{
\rvb&
{\begin{array}{c}
\rvy=\\f(\rvx)\end{array}}\ar[l]
&\rvx\ar[l]&\rva\ar[l]
}
\;\label{eq-dp-det-graph}
\eeq
where node $\rvy$ is deterministic with
$P(y|x) = \delta(y, f(x))$.
The data processing inequality
applied to graph Eq.(\ref{eq-dp-det-graph})
gives

\beq
H(f(\rvx):\rva) \leq H(\rvx:\rva)
\;,
\label{eq-h-fx-a}
\eeq
and

\beq
H(\rvb:\rvx) \leq H(\rvb:f(\rvx))
\;.
\label{eq-h-fx-b}
\eeq
Note that for any random
variable $\rvz$, one has

\beq
H(f(\rvz):\rvz)=H(f(\rvz)) - H(f(\rvz)|\rvz)=H(f(\rvz))
\;.\label{eq-info-fz}
\eeq
Combining Eqs.(\ref{eq-h-fx-a})
and (\ref{eq-info-fz}) yields\footnote{
What we really mean by
the limit $\rva\rarrow \rvx$
is that $P(x|a)=\delta_x^a$.
Taking $\rvb\rarrow\rvx$
in Eq.(\ref{eq-h-fx-b}) would not
work
because $\rvb$ and $\rvx$
are not adjacent to each other
whereas $\rva$ and $\rvx$ are.}

\beq
H(f(\rvx))=
H(f(\rvx):\rva)|_{\rva\rarrow\rvx}
\; \leq \;
H(\rvx:\rva)|_{\rva\rarrow\rvx}=
H(\rvx)
\;.
\eeq

Now let's try to find
quantum analogues to
the classical data
processing inequalities.
To do so, we will use
the following QB nets.

For
$j\geq 1$, let
$\beta_j = (\rvb_j,\rve_j)$. Define

\beq
\begin{array}{c}
\entrymodifiers={++[o][F-]}
\xymatrix{
*{}
&
\calg_j\ar[l]
&
*{}\ar[l]
}
\end{array}
=
\begin{array}{c}
\entrymodifiers={++[o][F-]}
\xymatrix{
*{}
&
\cancel{\rvb_j}\ar[l]
&
\rv{\beta}_j\ar[l]_>>{\delta}\ar[d]_>>{\delta}
&
*{}\ar[l]
\\
*{}
&
*{}
&
\cancel{\rve_j}
&
*{}
}
\end{array}
\;,
\eeq
and

\beq
\rho^{(j)}_{\rv{\beta}_j,\ldots,
\rv{\beta}_2,
\rv{\beta}_1, \rvb,\rva}
=
\sandb{
\entrymodifiers={++[o][F-]}
\xymatrix{
\calg_j&*{\;\dots\;}\ar[l]&\calg_2\ar[l]&
\calg_1\ar[l]&
\rvb\ar[l]&
\rva\ar[l]
}
}
\sandb{\hc}
\;.
\label{eq-qb-net-j-links}
\eeq
For example,

\beq
\rho^{(2)}_{\rv{\beta}_2,\rv{\beta}_1, \rvb,\rva}
=
\sandb{
\entrymodifiers={++[o][F-]}
\xymatrix{
\cancel{\rvb_2}&
\rv{\beta}_2\ar[l]_>>{\delta}\ar[d]_>>{\delta}&
\cancel{\rvb_1}\ar[l]&
\rv{\beta}_1\ar[l]_>>{\delta}\ar[d]_>>{\delta}&
\rvb\ar[l]&
\rva\ar[l]
\\
*{}&
\cancel{\rve_2}&
*{}&
\cancel{\rve_1}&
*{}&
*{}
}
}
\sandb{\hc}
\;,
\label{eq-qb-net-two-links}
\eeq

\beq
\rho^{(1)}_{\rv{\beta}_1, \rvb,\rva}=
{\rm erase}_{\rv{\beta}_2}\{
\rho^{(2)}_{\rv{\beta}_2,\rv{\beta}_1, \rvb,\rva}
\}
=
\sandb{
\entrymodifiers={++[o][F-]}
\xymatrix{
\cancel{\rvb_1}&
\rv{\beta}_1\ar[l]_>>{\delta}\ar[d]_>>{\delta}&
\rvb\ar[l]&
\rva\ar[l]
\\
*{}&
\cancel{\rve_1}&
*{}&
*{}
}
}
\sandb{\hc}
\;,
\eeq
and

\beq
\rho^{(0)}_{\rvb,\rva}=
{\rm erase}_{\rv{\beta}_1}\{
\rho^{(1)}_{\rv{\beta}_1, \rvb,\rva}
\}
=
\sandb{
\entrymodifiers={++[o][F-]}
\xymatrix{
\rvb&
\rva\ar[l]
}
}
\sandb{\hc}
\;.
\eeq
Note that the operations of
tracing versus erasing
a node from a density matrix
(and corresponding QB net)
are different. They can produce
different density matrices.

Let $\rvb_0=\rvb$.
For $j\geq 1$, assume the amplitude
$A(\beta_j|b_{j-1})$
comes from a
channel superoperator
$\calt_{\beta_j|b_{j-1}}$.
Hence, it must be an isometry.

Some
quantum data processing inequalities
refer to a single QB net,
whereas others refer to multiple ones.
The next two sections
address these two possibilities.

\subsubsection{Single-Graph Data Processing}
\begin{claim}
For ${\rho^{(2)}_{\rv{\beta}_{2},\rv{\beta}_{1},\rvb,\rva}}$ given by the QB net of
Eq.(\ref{eq-qb-net-two-links}),

\beq
S_{\rho^{(2)}_{\rv{b}_{2cl},\rva}}(
\rv{b}_{2cl}:\rva)
\stackrel{(a)}{\leq}
S_{\rho^{(2)}_{\rv{b}_{1cl},\rva}}(
\rv{b}_{1cl}:\rva)
\stackrel{(b)}{\leq}
S_{\rho^{(2)}_{\rvb_{cl},\rva}}(
\rvb_{cl}:\rva)
\;.
\eeq
(Note that the $\rve_j$ have
been traced over.)
\end{claim}
\proof
Inequality $(b)$
is just a special case of inequality $(a)$.
Inequality $(a)$ can be established
as follows.

\beqa
\lefteqn{
\atrho{
\rho^{(2)}_{\rv{b}_{2cl},\rv{b}_{1cl},\rva}
}
S(\rv{b}_1:\rva)
-
S(\rv{b}_{2}:\rva)
}\nonumber
\\
&=&
S(\rv{b}_1:\rva)-
[
S(\rv{b}_1,\rv{b}_2:\rva)
-
S(\rv{b}_1:\rva|\rv{b}_2)]
\label{eq-single-gr-dp-a}
\\
&=&
-S(\rv{b}_2:\rva|\rv{b}_1)
+
S(\rv{b}_1:\rva|\rv{b}_2)
\label{eq-single-gr-dp-b}
\\
&=&
S(\rv{b}_1:\rva|\rv{b}_2)
\label{eq-single-gr-dp-c}
\\
&\geq& 0
\label{eq-single-gr-dp-d}
\atrhoend
\;.
\eeqa
\begin{itemize}
\item[(\ref{eq-single-gr-dp-c}):] Follows because
$S(\rv{b}_2:\rva|\rv{b}_1)=0$
since
$\rv{b}_1$
is classical and at the middle of a
Markov chain.
See Claim \ref{cl-cond-mid-markov}.
\item[(\ref{eq-single-gr-dp-d}):] Follows
because CMI$\geq$ 0.
\end{itemize}
\mbox{\;}
\qed

\subsubsection{Multi-Graph Data Processing}

The following claim
was proven
by Schumacher and Nielsen
in Ref.\cite{schu-niel}.

\begin{claim}\label{cl-multi-gr-dp}
For ${\rho^{(j)}_{\rv{\beta}_j,\ldots,\rv{\beta}_{2},\rv{\beta}_{1},\rvb,\rva}}$ given by the QB net of
Eq.(\ref{eq-qb-net-j-links}),

\beq
S_{
\rho^{(3)}_{
\rvb_3,\cancel{\rvb_2},\cancel{b_1},\rva}
}(\rvb_3:\rva)
\stackrel{(a)}{\leq}
S_{
\rho^{(2)}_{
\rvb_2,\cancel{b_1},\rva}
}(\rvb_2:\rva)
\stackrel{(b)}{\leq}
S_{
\rho^{(1)}_{
\rvb_1,\rva}
}(\rvb_1:\rva)
\;.
\eeq
(Note that the $\rve_j$ have
been traced over.)\end{claim}
\proof

Inequalities $(a)$ and $(b)$
both follow from MRE
because
\beq
\rho^{(3)}_{
\rvb_3,\cancel{\rvb_2},\cancel{b_1},\rva}
=
\tr_{\rve_3}\circ
\calt_{\rv{\beta_3}|\rvb_2}
(\rho^{(2)}_{
\rvb_2,\cancel{b_1},\rva})
\;,
\eeq
and

\beq
\rho^{(2)}_{
\rvb_2,\cancel{b_1},\rva}
=
\tr_{\rve_2}\circ
\calt_{\rv{\beta_2}|\rvb_1}
(\rho^{(1)}_{
\rvb_1,\rva})
\;.
\eeq
\qed

\section{Hybrid Entropies With Both
Classical
and Quantum Random Variables}

\subsection{
Conditioning Entropy on a Classical Random Variable}

From the
definition
of $H(\rvb|\rva)$,
it's clear
that $H(\rvb|\rva)\geq 0$.
On the
other hand, $S(\rvb|\rva)$
can sometimes be negative.
One case where $S(\rvb|\rva)$
is guaranteed to
be non-negative
is when the random variable
being conditioned
on is classical.

\begin{claim}\label{cl-cond-on-class-rv}
For any $\rho_{\rva,\rvb}\in dm(\calh_{\rva,\rvb})$,
\beq
S_{\rho_{\rvb,\rva_{cl}}}
(\rvb|\rva_{cl})\geq
\max(0, S_{\rho_{\rvb,\rva}}(\rvb|\rva))
\;.
\eeq
\end{claim}
\proof
By MRE with $\calt = {\rm cl}_\rva$,

\beq
S(\rvb:\rva_{cl})
\leq
S(\rvb:\rva)
\;.
\eeq
But

\beq
S(\rvb:\rva_{cl})
=S(\rvb)-S(\rvb|\rva_{cl})
\;,
\eeq
and

\beq
S(\rvb:\rva)
=S(\rvb)-S(\rvb|\rva)
\;.
\eeq
Hence

\beq
S(\rvb|\rva_{cl})
\geq
S(\rvb|\rva)
\;.
\eeq

One can express $\rho_{\rvb, \rva_{cl}}$ as
\beq
\rho_{\rvb, \rva_{cl}}=
\sum_a
P(a)
\sandb{\ket{a}_\rva}
\rho_{\rvb|a}
\sandb{\hc}
\;
\eeq
where $\{P(a)\}_{\forall a}\in pd(S_\rva)$
and
$\rho_{\rvb|a}\in dm(\calh_\rvb)$
for all $a$. Therefore

\beqa
S(\rho_{\rvb, \rva_{cl}})
&=&
-\tr_\rvb
\sum_a P(a)\rho_{\rvb|a}
\ln\left(
P(a)\rho_{\rvb|a}\right)
\\
&=&
H\{P(a)\}_{\forall a} +
\sum_a P(a) S(\rho_{\rvb|a})
\;.
\eeqa
Hence,

\beq
S_{\rho_{\rvb, \rva_{cl}}}(\rvb|\rva_{cl})
=
S(\rho_{\rvb, \rva_{cl}})
-
S(\rho_{\rva_{cl}})=
\sum_a P(a) S(\rho_{\rvb|a})
\geq
0
\;.
\eeq
\qed
\subsection{Clone Random Variables}

We'll say two random variables are
clones of each other
if they are
 perfectly correlated.
Classical and quantum
clone random variables
behave very
differently as far as
entropy is concerned.
In this section, we
will show that
two classical clones
can be merged without changing the
entropy,
but not so for two quantum clones.

Consider the following CB net
\beq
\entrymodifiers={++[o][F-]}
\xymatrix{
\rva'&\rva\ar[l]\ar[r]&\rvb
}
\;,
\eeq
where

\beq
P(a'|a) = \delta_a^{a'}
\;.
\eeq
Since $P(a',a)=P(a) \delta_a^{a'}$,
one gets

\beq
H(\rva,\rva') = H(\rva)=H(\rva')
\;,
\eeq

\beq
H(\rva|\rva')= H(\rva'|\rva)=0
\;,
\eeq

\beq
H(\rva:\rva')= H(\rva)=H(\rva')
\;,
\eeq

\beq
H(\rvb, \rva, \rva') = H(\rvb, \rva)
=H(\rvb,\rva')
\;,
\eeq

\beq
H(\rvb, \rva|\rva')=
H(\rvb|\rva)=H(\rvb|\rva')
\;.
\eeq
All these
results can be described
by saying that
the classical clone
random variables
$\rva$ and $\rva'$
are interchangeable and
that often they can  be ``merged"
into a single random variable
without changing the entropy.

Quantum clone random variables,
on the other hand, cannot be merged
in general.
For example, for a general state
$\rho_{\rva,\rva'}$,
one has
$S(\rva,\rva')\neq S(\rva)$,
even if

\beq
\av{a,a'|\rho_{\rva,\rva'}|a,a'}
\varpropto \delta_a^{a'}
\;\label{eq-av-a-rho-a}
\eeq
for all $a,a'\in S_\rva$.
For example,
when

\beq
\rho_{\rva,\rva'}=
\sandb{
\frac{1}{\sqrt{N_\rva}}
\sum_{a}
\begin{array}{l}
\ket{a}_\rva\\
\ket{a}_{\rva'}
\end{array}
}
\sandb{\hc}
\;,
\eeq
Eq.(\ref{eq-av-a-rho-a}) is satisfied.
However,
$S(\rva,\rva')=0$ and
$S(\rva)=S(\rva')\neq 0$.
Hence,
 $S(\rva,\rva')\neq S(\rva)$.

Similarly, for a general state
$\rho_{\rvb,\rva,\rva'}$,
$S(\rvb,\rva,\rva')\neq S(\rvb,\rva)$.
For example,
when

\beq
\rho_{\rvb,\rva,\rva'}=
\sandb{\frac{1}{\sqrt{N_\rva N_\rvb}}
\sum_{a,b}
\begin{array}{l}
\ket{b}_\rvb\\
\ket{a}_\rva\\
\ket{a}_{\rva'}
\end{array}
}
\sandb{\hc}
\;,\label{eq-rho-baa-pure}
\eeq
Eq.(\ref{eq-av-a-rho-a}) is satisfied.
However,
$S(\rvb,\rva,\rva')=0$ and
$S(\rvb,\rva)=S(\rva')\neq 0$.
Hence,
$S(\rvb,\rva,\rva')\neq S(\rvb,\rva)$.

\begin{claim}
Suppose
\beq
\rho_{\rvb, \rva, \rva'}
=
\rho_{\rvb, \rva_{cl}, \rva'_{cl}}
=
\sum_a P(a)
\sandb{
\begin{array}{l}
\ket{a}_{\rva}\\
\ket{a}_{\rva'}
\end{array}
}
\rho_{\rvb|a}
\sandb{\hc}
\;
\eeq
where $\{P(a)\}_{\forall a}
\in pd(S_\rva)$
and
$\rho_{\rvb|a}\in dm(\calh_\rvb)$
for all $a$.
Then

\beq
S(\rvb,\rva,\rva')=
S(\rvb, \rva)= S(\rvb, \rva')
\;,
\eeq
and
\beq
S(\rvb,\rva|\rva')= S(\rvb| \rva)=
S(\rvb| \rva')
\;.
\eeq
\end{claim}
\proof

For any density matrix $\rho$
with no zero eigenvalues,
$\ln \rho$ can be
expressed as an infinite power
series
in powers of $\rho$:

\beq
\ln \rho=
\sum_{j=0}^{\infty} c_j \rho^j
\;,
\eeq
for some real numbers
$c_j$ that are independent of $\rho$.

Note that

\beqa
\tr_{\rva'}\rho^2_{\rvb, \rva, \rva'}
&=&
\tr_{\rva'}
\sum_a P^2(a)
\sandb{
\begin{array}{l}
\ket{a}_{\rva}\\
\ket{a}_{\rva'}
\end{array}
}
\rho^2_{\rvb|a}
\sandb{\hc}
\\
&=&
\sum_a P^2(a)
\sandb{
\ket{a}_{\rva}
}
\rho^2_{\rvb|a}
\sandb{\hc}
\\
&=&
\rho^2_{\rvb, \rva}
\;.
\eeqa
Thus, the operations of $\tr_{\rva'}$
and raising-to-a-power commute
when
acting on $\rho_{\rvb, \rva_{cl}, \rva'_{cl}}$.
(This is not the case for $\rho_{\rvb,\rva,\rva'}$
given by Eq.(\ref{eq-rho-baa-pure})).

Finally, note that

\beqa
S(\rvb,\rva,\rva')&=&
-\tr_{\rvb,\rva,\rva'}
\left(
\rho_{\rvb,\rva,\rva'}
\ln
\rho_{\rvb,\rva,\rva'}
\right)
\\
&=&
-\tr_{\rvb,\rva,\rva'}
\left(
\sum_{j=0}^\infty
c_j
\rho^{j+1}_{\rvb,\rva,\rva'}
\right)
\\
&=&
-\tr_{\rvb,\rva}
\left(
\sum_{j=0}^\infty
c_j
\rho^{j+1}_{\rvb,\rva}
\right)
\\
&=&
S(\rvb,\rva)=S(\rvb,\rva')
\;.
\eeqa
\qed

\subsection{Conditioning CMI On the
Middle of a Tri-node Markov-Like Chain}

We will refer to a
node with 2 incoming
arrows and no outgoing ones
as a collider.
Let's consider
all CB nets with 3 nodes
and 2 arrows. These can have
either one collider
or none.

The CB net with one collider
is
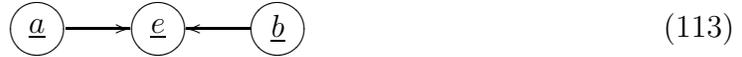
\beq
\entrymodifiers={++[o][F-]}
\xymatrix{
\rva\ar[r]&\rve&\rvb\ar[l]
}
\;
\eeq
For this net,
$P(a,b|e)\neq P(a|e) P(b|e)$
so
$H(\rva:\rvb|\rve)\neq 0$.

There
are 3 CB nets with no collider:
the
fan-out (a.k.a. broadcast, or fork) net,
and 2 Markov chains (in opposite directions):

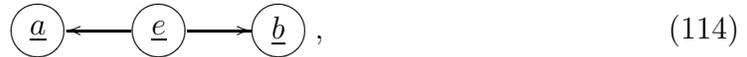
\beq
\entrymodifiers={++[o][F-]}
\xymatrix{
\rva&\rve\ar[l]\ar[r]&\rvb
}
\;,
\eeq

\beq
\entrymodifiers={++[o][F-]}
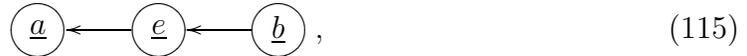
\xymatrix{
\rva&\rve\ar[l]&\rvb\ar[l]
}
\;,
\eeq

\beq
\entrymodifiers={++[o][F-]}
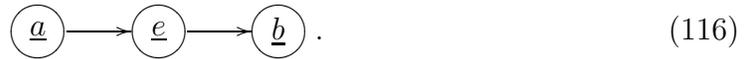
\xymatrix{
\rva\ar[r]&\rve\ar[r]&\rvb
}
\;.
\eeq
We will refer to these 3 graphs
as tri-node Markov-like chains.
For all 3 of these nets
$P(a,b|e)= P(a|e) P(b|e)$ so
$H(\rva:\rvb|\rve)= 0$.
In this case we say
$\rva$ and $\rvb$ are
conditionally
independent (of $\rve$).

\begin{claim}\label{cl-cond-mid-markov}
Let
\beq
\rho^{\rm fan-out}_{\rva,\rvb,\rve}=
\begin{array}{l}
\tr_{\rv{\alpha}_0,\rv{\eps}_0,\rv{\beta}_0}\\
\tr_{\rv{\alpha}_1,\rv{\eps}_1,\rv{\beta}_1}
\end{array}\sandb{
\entrymodifiers={++[o][F-]}
\xymatrix{
\rv{\alpha}_0\ar[d]&\rv{\eps}_0\ar[d]&\rv{\beta}_0\ar[d]\\
\rva\ar[d]&\rve\ar[l]\ar[d]\ar[r]&\rvb\ar[d]\\
\rv{\alpha}_1&\rv{\eps}_1&\rv{\beta}_1
}}\sandb{\hc}
\;,
\eeq
and

\beq
\rho^{Markov}_{\rva,\rvb,\rve}=
\begin{array}{l}
\tr_{\rv{\alpha}_0,\rv{\eps}_0,\rv{\beta}_0}\\
\tr_{\rv{\alpha}_1,\rv{\eps}_1,\rv{\beta}_1}
\end{array}
\sandb{
\entrymodifiers={++[o][F-]}
\xymatrix{
\rv{\alpha}_0\ar[d]&\rv{\eps}_0\ar[d]&\rv{\beta}_0\ar[d]\\
\rva\ar[d]&\rve\ar[l]\ar[d]&\rvb\ar[d]\ar[l]\\
\rv{\alpha}_1&\rv{\eps}_1&\rv{\beta}_1
}}\sandb{\hc}
\;.
\eeq
With $\rho_{\rva,\rvb,\rve}$
equal to either
$\rho_{\rva,\rvb,\rve}^{\rm fan-out}$
or
$\rho_{\rva,\rvb,\rve}^{\rm Markov}$,

\beq
S_{\rho_{\rva,\rvb,\rve_{cl}}}(\rva:\rvb|\rve_{cl})=0
\;.
\eeq
\end{claim}
\proof

At the end of this proof,
we will show that
for both of these QB nets,
$\rho_{\rva,\rvb,\rve_{cl}}$
can be expressed as

\beq
\rho_{\rva,\rvb,\rve_{cl}}=
\sum_e P(e)\sandb{\ket{e}_\rve}
\rho_{\rva|e}\;\rho_{\rvb|e}
\sandb{\hc}
\;,
\label{eq-rho-a-b-e}
\eeq
where $\{P(e)\}_{\forall e}\in pd(S_\rve)$,
and
 $\rho_{\rva|e}\in dm(\calh_\rva)$,
$\rho_{\rvb|e}\in dm(\calh_\rvb)$
for all $e\in S_\rve$.
Let's assume this for now. Then

\beqa
S(\rva,\rvb,\rve_{cl})&=&
-\tr_{\rva,\rvb}\sum_e\left\{
P(e)\rho_{\rva|e}\;\rho_{\rvb|e}
\ln\left(P(e)
\rho_{\rva|e}\;\rho_{\rvb|e}\right)
\right\}
\\
&=&
H\{P(e)\}_{\forall e}
+
\sum_e P(e)
[S(\rho_{\rva|e}) + S(\rho_{\rvb|e})]
\;.
\eeqa
Hence

\beq
S(\rva,\rvb|\rve_{cl})
=
\sum_e P(e)
[S(\rho_{\rva|e}) + S(\rho_{\rvb|e})]
\;.
\eeq
One can show in the same way that also

\beq
S(\rva|\rve_{cl})
=
\sum_e P(e)
S(\rho_{\rva|e})
\;,
\eeq
and

\beq
S(\rvb|\rve_{cl})
=
\sum_e P(e)
S(\rho_{\rvb|e})
\;.
\eeq
Thus

\beq
S(\rva:\rvb|\rve_{cl})
=
S(\rva|\rve_{cl})
+
S(\rvb|\rve_{cl})
-
S(\rva,\rvb|\rve_{cl})
=
0
\;.
\eeq

Now let's show
that $\rho_{\rva,\rvb,\rve_{cl}}$
has the form Eq.(\ref{eq-rho-a-b-e})
for both QB nets.

For the fan-out net,

\beq
\rho^{\rm fan-out}_{\rva,\rvb,\rve_{cl}}=
\sum_{\alpha_0, \eps_0, \beta_0}
\sum_{\alpha_1, \eps_1, \beta_1}
\sum_e
\sandb{
\begin{array}{r}
\sum_a A(a|e,\alpha_0)A(\alpha_1|a)A(\alpha_0)\ket{a}_\rva\\
\sum_b A(b|e,\beta_0)A(\beta_1|b)A(\beta_0)\ket{b}_\rvb\\
A(e|\eps_0)A(\eps_1|e)A(\eps_0)\ket{e}_\rve
\end{array}
}\sandb{\hc}
\;.
\eeq
Set

\beq
\rho_{\rva|e}
= C_{\rva|e}
\sum_{\alpha_0,\alpha_1} \sandb{\sum_a
A(a|e,\alpha_0)A(\alpha_1|a)A(\alpha_0)
\ket{a}_\rva}
\sandb{\hc}
\;,
\eeq
and

\beq
\rho_{\rvb|e}
= C_{\rvb|e}
\sum_{\beta_0,\beta_1} \sandb{\sum_b
A(b|e,\beta_0)A(\beta_1|b)A(\beta_0)
\ket{b}_\rvb}
\sandb{\hc}
\;.
\eeq
For $\rvx = \rva,\rvb$,
the constant $C_{\rvx|e}$
depends
on $e$ and is defined so that
$\tr_\rvx \rho_{\rvx|e}=1$.

For the Markov chain net,
\beq
\rho^{\rm Markov}_{\rva,\rvb,\rve_{cl}}=
\sum_{\alpha_0, \eps_0, \beta_0}
\sum_{\alpha_1, \eps_1, \beta_1}
\sum_e
\sandb{
\begin{array}{r}
\sum_a A(a|e,\alpha_0)A(\alpha_1|a)A(\alpha_0)\ket{a}_\rva\\
\sum_b A(b|\beta_0)A(\beta_1|b)A(\beta_0)\ket{b}_\rvb\\
A(e|b,\eps_0)A(\eps_1|e)A(\eps_0)\ket{e}_\rve
\end{array}
}\sandb{\hc}
\;.
\eeq
Set

\beq
\rho_{\rva|e}
= C_{\rva|e}
\sum_{\alpha_0,\alpha_1} \sandb{\sum_a
A(a|e,\alpha_0)A(\alpha_1|a)A(\alpha_0)
\ket{a}_\rva}
\sandb{\hc}
\;,
\eeq
and

\beq
\rho_{\rvb|e}
= C_{\rvb|e}
\sum_{\eps_0,\eps_1}
\sum_{\beta_0,\beta_1} \sandb{\sum_b
\begin{array}{l}
A(b|\beta_0)A(\beta_1|b)A(\beta_0)\\
A(e|b,\eps_0)A(\eps_1|e)A(\eps_0)
\end{array}
\ket{b}_\rvb}
\sandb{\hc}
\;,
\eeq
where again,
$C_{\rva|e}$ and
$C_{\rvb|e}$ are
defined so that
the density matrices
$\rho_{\rva|e}$ and $\rho_{\rvb|e}$
have unit trace.

For both graphs, if we define
\beq
P(e)= \tr_{\rva,\rvb}\av{e|\rho_{\rva,\rvb,\rve}|e}
\;,
\eeq
then Eq.(\ref{eq-rho-a-b-e}) is
satisfied.
\qed

\subsection{Tracing the Output of an Isometry}
This section
will mention an observation
that is pretty trivial, but
arises frequently so it is
worth pointing out explicitly.

Consider the following  density matrix
\beq
\rho_{\rvc,\rvb,\rva}=
\sandb{
\entrymodifiers={++[o][F-]}
\xymatrix{
\rvc&\rvb\ar[l]&\rva\ar[l]
}
}
\sandb{\hc}
=
\sandb{
\begin{array}{r}
A(c|b)\ket{c}_\rvc\\
\sum_{a,b,c}
A(b|a)\ket{b}_\rvb\\
A(a)\ket{a}_\rva
\end{array}
}
\sandb{\hc}
\;.
\eeq
{\it Assume that $A(c|b)$ is an isometry.}
Then

\beq
\tr_\rvc
\rho_{\rvc,\rvb,\rva}
=
\sum_b
\sandb{
\begin{array}{l}
A(b|a)\ket{b}_\rvb\\
A(a)\ket{a}_\rva
\end{array}
}
\sandb{\hc}
=
\rho_{\rvb_{cl},\rva}
\;,
\eeq
and

\beq
\atrho{
\rho_{\rvc,\rvb,\rva}}
S(\rvb,\rva)=S(\rvb_{cl},\rva)
\atrhoend
\;.
\eeq
Thus, we observe
that tracing over all the {\it output}
indices of an isometry amplitude
embedded within a density matrix converts
the {\it inputs} of that isometry
amplitude into classical
random variables.

Next consider
the following density matrix,

\beq
\rho_{\rvc,\cancel{\rvb},\rva}=
\sandb{
\entrymodifiers={++[o][F-]}
\xymatrix{
\rvc&\cancel{\rvb}\ar[l]&\rva\ar[l]
}
}
\sandb{\hc}
=
\sandb{
\begin{array}{r}
A(c|b)\ket{c}_\rvc\\
\sum_{a,b,c}
A(b|a)A(a)\ket{a}_\rva
\end{array}
}
\sandb{\hc}
\;.
\eeq
{\it Assume that both $A(c|b)$ and $A(b|a)$
are isometries}.
Then

\beq
\tr_\rvc
\rho_{\rvc,\cancel{\rvb},\rva}
=
\sum_a
\sandb{
\begin{array}{l}
A(a)\ket{a}_\rva
\end{array}
}
\sandb{\hc}
=
\rho_{\rva_{cl}}
\;,
\eeq
and

\beq
\atrho{
\rho_{\rvc,\cancel{\rvb},\rva}}
S(\rva)=S(\rva_{cl})
\atrhoend
\;.
\eeq
Thus, we observe
that two isometries
joined by slashed variables
behave as if they
were just one isometry.

\subsection{Holevo Information}

Suppose $\{P(x)\}_{\forall x}\in pd(S_\rvx)$
and $\rho_{\rvq|x}\in dm(\calh_\rvq)$
for all $x$. Set

\beq
\rho_\rvq=
E_x \rho_{\rvq|x}
=
\sum_x
P(x)
\rho_{\rvq|x}
=
\sum_x
\sandb{\sqrt{P(x)}}
\rho_{\rvq|x}
\sandb{\hc}
\;.\label{eq-rhoq-px-rhoqx}
\eeq
Then the Holevo information
for the ensemble
$\{P(x),\rho_{\rvq|x}\}_{\forall x}$
is defined as

\beq
Hol\{P(x),\rho_{\rvq|x}\}_{\forall x}= S(E_x \rho_{\rvq|x})-
E_x S(\rho_{\rvq|x})=[S,E_x]\rho_{\rvq|x}
\;.
\eeq

\begin{claim}\label{claim-hol-mi}
Let
\beq
\rho_{\rvq,\rvx}=
\rho_{\rvq,\rvx_{cl}}=
\sum_x P(x)
\sandb{\ket{x}_\rvx}\rho_{\rvq|x}
\sandb{\hc}
\;,
\eeq
where $\{P(x)\}_{\forall x}\in pd(S_\rvx)$
and
$\rho_{\rvq|x}\in dm(\calh_\rvq)$ for all $x$.
Then
\beq
Hol\{P(x),\rho_{\rvq|x}\}_{\forall x} = S_{\rho_{\rvq, \rvx_{cl}}}(
\rvq : \rvx_{cl})
\;
\eeq
Thus,
the Holevo
information is a
MI with one of the
two random variables classical.
\end{claim}
\proof

\beqa
S_{\rho_{\rvq, \rvx_{cl}}}(\rvq,\rvx_{cl})
&=&
-\tr_\rvq
\sum_x \left\{
P(x)
\rho_{\rvq|x}
\ln\left(P(x)\rho_{\rvq|x}\right)
\right\}
\\
&=&
H\{P(x)\}_{\forall x}
+
E_x S(\rho_{\rvq|x})
\;.
\eeqa
Hence

\beqa
\atrho{\rho_{\rvq, \rvx_{cl}}}
S(\rvq:\rvx_{cl})
&=&
S(\rvq)
-S(\rvq|\rvx_{cl})
\\
&=&
S(\rvq)-
E_x S(\rho_{\rvq|x})
\\
&=&
S(E_x \rho_{\rvq|x})-
E_x S(\rho_{\rvq|x})
\atrhoend
\;.
\eeqa
\qed

\section{Holevo Bound}

In this section
we prove the so called
Holevo Bound, which
is an upper bound
 on the
accessible information. The accessible
information
is a figure of merit of
a quantum ensemble.
The upper
bound
is given by
the Holevo
information.
The proof of
the Holevo Bound\footnote{
The proof given here
of Holevo's original result (Ref.\cite{hol})
is very similar to the one
first given by Schumacher and Westmoreland
in Ref.\cite{schu-west}.}
that we give next, it
utilizes and therefore illustrates
many of
the concepts and inequalities
that were introduced earlier in
this paper.

Consider a density
matrix $\rho_\rvq$ expressible
in the form Eq.(\ref{eq-rhoq-px-rhoqx}).
It is useful
to re-express
$\rho_\rvq$
using
the eigenvalue decompositions
of the density matrices
$\rho_{\rvq|x}$.
For some $\rv{Q}$
with $S_{\rv{Q}}=S_\rvq$,
suppose the eigenvalue
decompositions of the
$\rho_{\rvq|x}$ are
given by

\beq
\rho_{\rvq|x}=
\sum_Q \lam_{Q|x}
\sandb{
\ket{\lam_{Q|x}
}_\rvq}
\sandb{\hc}
\;
\eeq
for all $x$.
Define

\beq
A(x) = \sqrt{P(x)}
\;,
\eeq

\beq
A(Q|x)=
\sqrt{\lam_{Q|x}}
\;,
\eeq
and

\beq
A(q|Q,x)=
\av{q|\lam_{Q|x}}
\;.
\eeq
Then

\beq
\rho_\rvq=
\sum_{x,Q}
\sandb{\sum_q A(q|Q,x)A(Q|x)A(x)\ket{q}_\rvq}
\sandb{\hc}
\;.
\eeq

It is useful to
find a purification of $\rho_\rvq$;
that is, a pure state
$\rho_{\rvq,\rvr}$
such that $\rho_\rvq =
\tr_{\rvr}(\rho_{\rvq,\rvr})$.
One possible purification of
$\rho_\rvq$ is given by

\beq
\rho_{\rvq,\rv{Q},\rvx_{cl}}
=
\sum_x
\sandbr{
\sum_{q,Q}
A(q|Q,x)\ket{q}_\rvq\\
A(Q|x)\ket{Q}_{\rv{Q}}\\
A(x)\ket{x}_\rvx
}
\sandb{\hc}
=
\sandb{
\entrymodifiers={++[o][F-]}
\xymatrix{
\rvq&*{}&\rvx_{cl}\ar[ll]\ar[dl]\\
*{}&\rv{Q}\ar[ul]&*{}
}
}
\sandb{\hc}
\;\label{eq-qb-net-rhoq}
\eeq
with $\rvr=(\rv{Q},\rvx_{cl})$.

Let $S_{\rvq_j}=S_{\rv{Q}}=S_\rvq$, and
$S_{\rvy_j}=S_\rvy$
for $j=1,2,3$. Suppose
$\rho_{\rvq_1}\in pd(\calh_{\rvq})$
is defined by
Eq.(\ref{eq-rhoq-px-rhoqx})
with $\rvq$ replaced by $\rvq_1$.
Suppose $\rho_{\rvq_1}$
is transformed to
$\rho_{\rvq_2}'\in dm(\calh_\rvq)$
by a quantum
channel with
Krauss operators $\{K_y\}_{\forall y}$. Thus

\beq
\rho_{\rvq_2}'=
\sum_y
K_y \rho_{\rvq_1}
K_y^\dagger
\;.
\eeq
As explained
in Ref.\cite{Tuc-mixology},
the Krauss operators $\{K_y\}_{\forall y}$
can be extended to a
unitary matrix $U_{\rvq,\rvy}$.
Let

\beq
\av{q_2|K_y|q_1}
=
\begin{array}{rcl}
\bra{q_2}_{\rvq_2}& & \ket{q_1}_{\rvq_1}\\
&U_{\rvq,\rvy}&\\
\bra{y}_{\rvy_2}& & \ket{0}_{\rvy_1}
\end{array}
\;
\eeq
for all $q_1,q_2\in S_\rvq$
and $y\in S_\rvy$.
Now we can define

\beq
R_{\rvq_2,\rvy_2,\rvx_{cl}}
=
\tr_{\rv{Q}}
\sandb{
\entrymodifiers={+++[o][F-]}
\xymatrix{
\cancel{\rvq_3}&*{}&
\cancel{\rvq_1}\ar[dl]&*{}&\rvx_{cl}\ar[dl]\ar[ll]\\
*{}&\rvq_2,\rvy_2\ar[ul]_>>{\delta}\ar[dl]_>>{\delta}&*{}&\rv{Q}\ar[ul]&*{}\\
\cancel{\rvy_3}&*{}&
\cancel{\rvy_1}\ar[ul]_<<{0}&*{}&*{}
}
}
\sandb{\hc}
\;,\label{eq-qb-net-hol}
\eeq
where

\beq
A(q_2,y_2|q_1, y_1=0)
=
\begin{array}{rcl}
\bra{q_2}_{\rvq_2}& & \ket{q_1}_{\rvq_1}\\
&U_{\rvq,\rvy}&\\
\bra{y_2}_{\rvy_2}& & \ket{0}_{\rvy_1}
\end{array}
\;
\eeq
for all $q_1,q_2\in S_\rvq$
and $y_2\in S_\rvy$.

Note that
$R_{\rvq_2,\rvy_2,\rvx_{cl}}$
satisfies
$R_{\rvq_2}=\rho'_{\rvq_2}$.

\begin{claim}\label{cl-hol}
If
$\rho_{\rvq_1,\rv{Q},\rvx_{cl}}$
is the QB net of Eq.(\ref{eq-qb-net-rhoq})
with $\rvq$ replaced by $\rvq_1$,
and
$R_{\rvq_2,\rvy_2,\rvx_{cl}}$
is the QB net of Eq.(\ref{eq-qb-net-hol}), then
\beq
S_{R_{\rvy_2,\rvx_{cl}}}(\rvy_2:\rvx_{cl})
\leq
Hol\{P(x), \rho_{\rvq_1|x}\}_{\forall x}
\;.
\eeq
\end{claim}
\proof
\beqa
S_{R_{\rvy_2,\rvx_{cl}}}(\rvy_2:\rvx_{cl})
&\leq&
S_{R_{\rvq_2,\rvy_2,\rvx_{cl}}}(\rvq_2,\rvy_2:\rvx_{cl})
\label{eq-hol-a}
\\
&\leq&
S_{\rho_{\rvq_1,\rvx_{cl}}}(\rvq_1:\rvx_{cl})
\label{eq-hol-b}
\\
&=&
Hol\{P(x), \rho_{\rvq_1|x}\}_{\forall x}
\label{eq-hol-c}
\;.
\eeqa

\begin{itemize}
\item[(\ref{eq-hol-a}):]
Follows because of MRE with $\calt=\tr_{\rvq_2}$.
\item[(\ref{eq-hol-b}):]
Follows from the multi-graph data processing
inequalities.
\item[(\ref{eq-hol-c}):] Follows
from Claim \ref{claim-hol-mi}.
\end{itemize}\mbox{\;}
\qed

Define the accessible information $Acc$
of the ensemble
$\{P(x), \rho_{\rvq_1|x}\}_{\forall x}$
and any channel with Krauss operators
$\{K_y\}_{\forall y}$
by
\beq
Acc\{P(x), \rho_{\rvq_1|x}\}_{\forall x}
=
\max_{\{K_y\}_{\forall y}}S_{R_{\rvy_2,\rvx_{cl}}}(\rvy_2:\rvx_{cl})
\;.
\eeq
Claim \ref{cl-hol} implies that

\beq
Acc\{P(x), \rho_{\rvq_1|x}\}_{\forall x}
\leq
Hol\{P(x), \rho_{\rvq_1|x}\}_{\forall x}
\;.
\eeq

\appendix
\section{Appendix: Schmidt Decomposition}
\label{app-s-decomp}
In this appendix, we define
the Schmidt decomposition of any
bi-partite pure state.

Consider any pure state $\ket{\psi}_{\rva,\rvb}\in
\calh_{\rva,\rvb}$. It can be expressed as

\beq
\ket{\psi}_{\rva,\rvb}=
\sandbr{
\sum_{a,b}
A(a,b)
\ket{a}_\rva
\\
\ket{b}_\rvb
}
\;.
\eeq
Assume $S_\rva\supset S_\rvb$.
Thus, $N_\rva\geq N_\rvb$.
$A(a,b)$ can be thought of as an
$N_\rva\times N_\rvb$
matrix. Let its
singular value decomposition be

\beq
A(a,b)=
\sum_{a_1\in S_\rva}
\sum_{b_1\in S_\rvb}
U(a,a_1)
\sqrt{P(b_1)}
\theta(a_1= b_1)
V^\dag
(b_1,b)
\;
\eeq
for all $a\in S_\rva$,
$b\in S_\rvb$,
where $U$ and $V$ are
unitary matrices.
Then we can express $\ket{\psi}_{\rva,\rvb}$
as

\beq
\ket{\psi}_{\rva,\rvb}=
\sandbr{
\sum_{b_1\in S_\rvb}
\sqrt{P(b_1)}
\ket{b_1}_\rva'
\\
\ket{b_1}_\rvb'
}
\;,
\label{eq-s-decomp}
\eeq
where

\beq
\ket{a_1}_\rva'=
\sum_{a\in S_\rva} U(a,a_1)\ket{a}_\rva
\;,
\eeq
for all $a_1\in S_\rva$ and

\beq
\ket{b_1}_\rvb'=
\sum_{b\in S_\rvb} V^*(b,b_1)\ket{b}_\rvb
\;
\eeq
for all $b_1\in S_\rvb$.
Eq.(\ref{eq-s-decomp})
is called
the Schmidt Decomposition
of $\ket{\psi}_{\rva,\rvb}$.

\begin{claim}
\label{claim-sa-is-sb}
If $\rho_{\rva,\rvb}\in dm(\calh_{\rva,\rvb})$
is pure, then

\beq
S(\rva)=S(\rvb)
\;.
\eeq
\end{claim}
\proof

Let
\beq
\rho_{\rva,\rvb}=
\sandb{
\ket{\psi}_{\rva,\rvb}
}
\sandb{\hc}
\;.
\eeq
If we express
$\ket{\psi}_{\rva,\rvb}$
as in Eq.(\ref{eq-s-decomp}), then

\beq
S(\rva)=
H\{P(b_1)\}_{\forall b_1\in S_\rvb}=
S(\rvb)
\;.
\eeq
\qed

\section{Appendix: Partial Entropies of
Pure Multi-Partite State}\label{app-partial-ents}

In this appendix,
we state some
consequences of Claim \ref{claim-sa-is-sb}
for the partial entropies
of pure multi-partite states.

Let $\rva_J = (\rva_j)_{j\in J}$
for any $J\subset Z_{1,N}$.

\begin{claim}\label{cl-sj-sjc}
Suppose
$J$ is
a nonempty subset of $\zn$ and
$J^c=Z_{1,N}-J$. If
$\rho_{\rva_\zn}$ is pure, then

\beq
\left\{
\begin{array}{l}
S(\rva_{\zn})=0,
\\
S(\rva_{J})=S(\rva_{J^c})
\end{array}
\right.
\;.
\eeq
For example, for $N=4$, this means

\beq
\left\{
\begin{array}{l}
S(\rva_1,\rva_2, \rva_3, \rva_4)=0,
\\
S(\rva_1,\rva_2,\rva_3) = S( \rva_4)
\mbox{ and permutations,}
\\
S(\rva_1,\rva_2) = S( \rva_3, \rva_4)
\mbox{ and permutations}
\end{array}
\right.
\;.
\eeq
\end{claim}
\proof
This is just a
generalization of Claim \ref{claim-sa-is-sb}.
\qed

\begin{claim}
Suppose $I,J$
are nonempty, disjoint subsets of $\zn$
such that $I\cup J = \zn$.
If $\rho_{\rva_\zn}$
is a pure state, then

\begin{subequations}
\beq
S(\rva_{I}|\rva_{J})=-S(\rva_{I})=-S(\rva_{J})
\;
\eeq

\beq
S(\rva_{I}:\rva_{J}) = 2 S(\rva_{I})=2 S(\rva_{J})
\;
\eeq
\end{subequations}
\end{claim}
\proof
Obvious.
\qed

\begin{claim}
Suppose $I,J,K$
are nonempty, disjoint subsets of $\zn$
such that $I\cup J\cup K = \zn$.
If $\rho_{\rva_\zn}$
is a pure state, then

\begin{subequations}
\beq
S(\rva_{J}|\rva_{I}) = S(\rva_{K}) -S(\rva_{I})
\;
\eeq

\beq
S(\rva_{J}|\rva_{I}) = -S(\rva_{J}|\rva_{K})
\;
\eeq

\beq
S(\rva_{I}:\rva_{J})
=
S(\rva_{I})+S(\rva_{J})-S(\rva_{K})
\;
\eeq

\beq
S(\rva_{I}:\rva_{J}|\rva_{K})= S(\rva_{I}:\rva_{J})
\;
\eeq
\end{subequations}
\end{claim}
\proof
Obvious.
\qed

\begin{claim}
Suppose $I,J,K,L$
are nonempty, disjoint subsets of $\zn$
such that $I\cup J\cup K\cup L = \zn$.
If $\rho_{\rva_\zn}$
is a pure state, then
\begin{subequations}
\beq
S(\rva_{I}:\rva_{J}|\rva_{K}) =
S(\rva_{I}|\rva_{K})-
S(\rva_{I}|\rva_{L})
\;
\eeq
\beq
S(\rva_{I}:\rva_{J}|\rva_{K}) =
-S(\rva_{I}:\rva_{J}|\rva_{L})
\;
\eeq
\end{subequations}
\end{claim}
\proof
Obvious.
\qed

\section{Appendix: RUM of Pure States}
In this appendix,
I describe
what I call
the RUM (Roots of Unity Model) of pure states.
The model only works
for pure states, and even
for those there is no guarantee
that it will always give the
right answer. That's why I
call it a model.

One famous physics ``model" is
 the Bohr model of
the Hydrogen atom. The Bohr model
 gives some nice intuition
about what is going on, plus it
predicts some (not all)
of the features of the Hydrogen
spectrum.

The RUM of pure states gives some
insight into why quantum
conditional entropies $S(\rvb|\rva)$
can be negative unlike classical
conditional entropies $H(\rvb|\rva)$
which are always non-negative.
It also gives some
insight into the identities
presented in Appendix \ref{app-partial-ents}
for the partial entropies of multi-partite states.
It ``explains" such identities
as being a consequence of
the high degree of symmetry
of pure multi-partite states.

Consider an $N$-partite
pure state described
by $N$ random variables
$\rva_1,\rva_2,\ldots,\rva_N$.
We {\it redefine} the
random variables $\rva_j$
so that they equal the $N$'th
roots of unity:

\beq
\rva_j = \exp(i \frac{2\pi(j-1)}{N})
\;
\eeq
for $j\in \zn$. Let
 $J$ be any nonempty subset of
$Z_{1,N}$.
Let $\sum \rva_{J} =
\sum_{j\in J} \rva_j$.
We {\it redefine} the entropy of the
$N$-partite state as follows

\beq
S(\rva_{J}) = \left|\sum \rva_{J}\right|
\;.
\eeq
Note that the various
subsystems $\rva_j$
contribute to this entropy in
a coherent sum,
instead of the incoherent
sums that we usually find
when dealing with classical entropy.

Note that

\beq
\sum \rva_J = - \sum \rva_{J^c}
\;
\eeq
so

\beq
S(\rva_{J}) = S(\rva_{J^c})
\;.
\eeq
This identity
was obtained in the exact case
too, in Claim \ref{cl-sj-sjc}.

Let $J,K$ be two nonempty disjoint
subsets of $Z_{1,N}$.
In this model
\beq
S(\rva_{K}|\rva_{J})
= S(\rva_{K},\rva_{J})-S(\rva_{J})=
\left|\sum\rva_{K\cup J}\right|
-\left|\sum\rva_{J}\right|
\;,
\eeq
which clearly
can be negative.

From the triangle inequalities

\beq
\left||\sum\rva_{J}|-|\sum\rva_{K}|
\right|
\leq
|\sum\rva_{J\cup K}|
\leq
|\sum\rva_{J}|+|\sum\rva_{K}|
\;.
\eeq
This can be re-written as

\beq
|S(\rva_{J})-S(\rva_{K})|
\leq
S(\rva_{J}, \rva_{K})
\leq
S(\rva_{J}) + S(\rva_{K})
\;.
\eeq
We recognize this as
the Araki-Lieb inequality and
subadditivity of the joint entropy.

\end{document}